\documentclass[10pt,letterpaper,twocolumn]{article} %% two column, final layout
\usepackage{ol}
\usepackage{hyperref}
\usepackage{amsmath,bm}
\usepackage{graphics}
\begin{document}
\twocolumn[ %% activate for two-column option
\title{Input-output relations for a 3-port grating coupled Fabry-Perot cavity}
\author{A. Bunkowski, O. Burmeister, K. Danzmann, and R. Schnabel}
\address{Max-Planck-Institut f\"ur Gravitationsphysik
(Albert-Einstein-Institut), and Institut f\"ur Atom- und
Molek\"ulphysik, \\Universit\"at Hannover, Callinstr. 38, 30167
Hannover, Germany}
\begin{abstract}
We analyze an optical 3-port reflection grating by means of a
scattering matrix formalism. Amplitude and phase relations between
the 3 ports, i.e. the 3 orders of diffraction are derived. Such a
grating can be used as an all-reflective, low-loss coupler to
Fabry-Perot cavities. We derive the input output relations of a
3-port grating coupled cavity and find distinct properties not
present in 2-port coupled cavities. The cavity relations further
reveal that the 3-port coupler can be designed such that the
additional cavity port interferes destructively. In this case the
all-reflective, low-loss, single-ended Fabry-Perot cavity becomes
equivalent to a standard transmissive, 2-port coupled cavity.
\end{abstract}
\ocis{050.1950, 120.3180, 230.1360.}
] %% activate for two-column option
\noindent In a recent experiment a 3-port reflection grating
coupled Fabry-Perot cavity with high Finesse was demonstrated
\cite{BBBDSCKT04}. The experiment was motivated by the idea that a
3-port reflection grating should be able to provide two important
features for advanced interferometry:  low  overall optical loss
and no light transmission through optical
substrates~\cite{Drever96}.
In advanced interferometers, such as for gravitational wave
detectors, these couplers will be crucial for achieving the
optimal combination of extremely high power laser fields,
materials of high mechanical quality factors for suspended optics
and cryogenic temperatures to reduce optics and suspension thermal
noise.~\cite{Rowan}
Previously, a different concept for all-reflective linear
Fabry-Perot cavities based on a two-port reflection grating was
experimentally demonstrated \cite{Sun97}. In this approach the
reflection grating was used in a 1st order Littrow mount where the
input-output relations of the cavity are analogous to those of a
conventional cavity with transmissive mirrors. The major
disadvantage of this concept is, however, that it relies on high
1st order diffraction efficiency requiring deep grating structures
that are associated with high scattering losses. Contrary to that,
the concept demonstrated in \cite{BBBDSCKT04} used a 2nd order
Littrow mount and relies on low 1st order diffraction efficiency
which can be achieved by very shallow grating structures with
smaller scattering losses. The latter approach is therefore better
suited for low-loss couplers to high-finesse cavities, a stringent
requirement in high-power laser interferometry. A grating used in
2nd order Littrow mount, however, has 3 coupled ports in contrast
to mirrors where one input port is only coupled to 2 output ports.
Knowledge of the phase relations of the three ports is essential
for the derivation of the input-output relations of the cavity.

\begin{figure}[h]\centerline{\scalebox{1}{\includegraphics{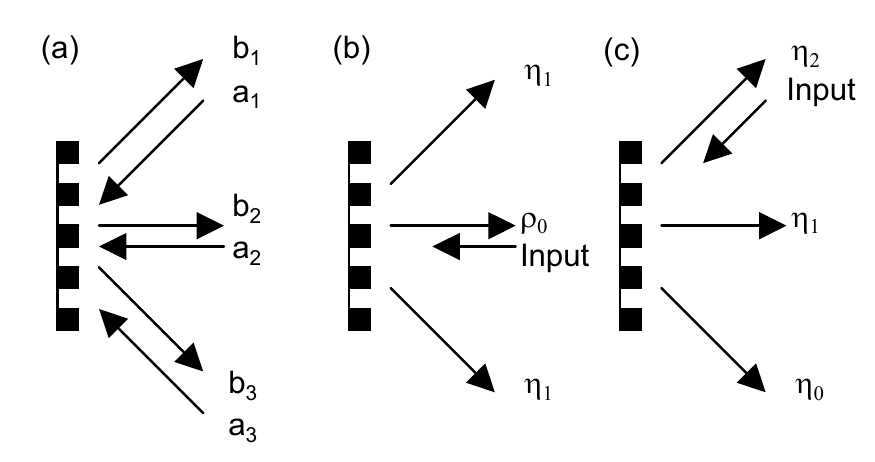}}}
  \caption{A 3-port reflection grating: (a) labelling of the input
  and output ports, (b) amplitudes of reflection coefficients for
  normal incidence, (c) for 2nd order Littrow incidence.}
   \label{fig:threeports}
\end{figure}

In this letter we derive the amplitude and phase relations of an
optical 3-port device by means of the scattering matrix formalism.
We restrict ourselves to a symmetric coupling between port 2 and
the other two ports 1 and 3 described by $\eta_1$, see
Fig.~\ref{fig:threeports}. Generally, optical devices like mirrors
and beam splitters can be described by a complex valued $n\times
n$ scattering matrix $\bm{S}$\cite{Siegman}, where  $n$ input
ports are represented by a vector $\bm{a}$ with the components
$a_i$ which are the complex amplitudes of the incoming waves at
the $i$th port. The outgoing amplitudes $b_i$ are represented by
the vector $\bm{b}$. The coupling of input and output ports is
given by the following equation
\begin{equation}\label{eq:s_matrix}
 \bm{b}= \bm{S}\times\bm{a}\,.
\end{equation}
For a loss-less device $\mbox{\boldmath $S$}$ must be unitary.
Reciprocity of the device demands $|S_{ij}|\equiv|S_{ji}|$, where
$S_{ij}$ denotes an element of the matrix $\mbox{\boldmath $S$}$.
The magnitudes of the scattering coefficients are unique for a
given device. The phase angles of the matrix elements however can
be changed by choosing different reference planes in the various
input and output arms. One can therefore derive different
scattering matrices for the same device. Nevertheless, certain
phase relationships between the different coefficients must be
maintained. Transmissive mirrors are commonly used to couple light
into Fabry-Perot cavities. The input output relations of such
cavities are well understood. Essential for their derivation is
the knowledge of the phase relations at the mirrors for the
reflected and transmitted beams. A conventional two-coupled-port
mirror with amplitude reflectance $\rho$ and transmittance $\tau$
for example is generally described by
\begin{equation}
\bm{S}_{2p}=
\left(%
\begin{array}{cc} \label{eq:bs}
  \rho & \tau \\
  \tau & -\rho \\
\end{array}%
\right) \quad \textrm{or}\quad \bm{S}_{2p}=
\left(%
\begin{array}{cc}
  \rho & i\tau \\
  i\tau & \rho \\
\end{array}%
\right).
\end{equation}
Using either one of these matrices one can derive the amplitude
reflectance $r_\mathrm{{FP}}$ and transmittance $t_{\mathrm{FP}}$
%and the intra-cavity amplitude $c_{\mathrm{FP}}$
of a cavity consisting of 2 partially transmitting mirrors of
reflectivities $\rho_0, \rho_1$. The length of the cavity is
expressed by the tuning parameter $\phi=\omega L/c$, where
$\omega$ is the angular frequency of the light and $c$ the speed
of light, thus one obtains
\begin{eqnarray}
r_{\mathrm{FP}}&=& [\rho_0 -\rho_1\exp(2i\phi)]d\,,\\
t_{\mathrm{FP}}&=& -\tau_0\tau_1 \exp(-i\phi) d
\end{eqnarray}
where $\rho_{0,1}$ and $\tau_{0,1}$ denote the reflectance and
transmittance of the two cavity mirrors, respectively, and we have
introduced the resonance factor
\begin{equation}\label{eq:d}
d = [1-\rho_0\rho_1\exp(2i\phi)]^{-1}\,.
\end{equation}
The power gain $g_{\mathrm{FP}}$ inside the cavity is given by
\begin{equation}\label{eq:gain}
g_{\mathrm{FP}}=|\tau_0 d|^2.
\end{equation}

The 3-port coupler used in reference\,\cite{BBBDSCKT04} can be
represented by the following scattering matrix
\begin{equation}
\label{eq:S} \bm{S}_{3p}\!=\!
\left(%
\begin{array}{lcr}
  \eta_2\exp(i\phi_{2}) \!&\! \eta_1\exp(i\phi_{1})   \!&\! \eta_0\exp(i\phi_0)      \!\\
  \eta_1\exp(i\phi_{1}) \!&\! \rho_0\exp(i\phi_0)     \!&\! \eta_1\exp({i\phi_{1}}) \!\\
  \eta_0\exp(i\phi_0)   \!&\! \eta_1\exp(i\phi_{1})   \!&\! \eta_2\exp(i\phi_{2})    \!\\
\end{array}%
\right).
\end{equation}

As stated above the grating is assumed to be symmetrical with
respect to the grating normal. The grating period and wavelength
of light is chosen in such a way that for normal incidence only
the 0th and 1st order diffraction are present. The magnitudes of
their amplitude reflection coefficients are denoted with $\rho_0$
and $\eta_1$ respectively. For incidence at the 2nd order Littrow
angle the 0th, 1st, and 2nd  diffraction orders are present with
the magnitudes of the reflection coefficients $\eta_0, \eta_1$ and
$\eta_2$ as depicted in Fig.~\ref{fig:threeports}. From the
unitarity condition of $\bm{S}$ we find the energy conservation
law
\begin{eqnarray}
\label{eq:energycon}
\rho_0^2+2\eta_1^2&\!=\!&1\,,\\
\eta_0^2+\eta_1^2+\eta_2^2&\!=\!&1\,.
\end{eqnarray}
We denote  the phase shift associated with the 0th, 1st, and 2nd
diffraction orders with $\phi_0, \phi_1$ and $\phi_2$,
respectively. As for mirrors the values for the phases are not
unique. Reflection from a mirror is equivalent to 0th order
diffraction of a grating. In analogy to the right matrix of
equation~(\ref{eq:bs}) we demand no phase shift for 0th order
diffraction and therefore set $\phi_0=0$. From the unitarity
requirement of $\bm{S}$ the remaining phases can be calculated
yielding the following possible set of phases
\begin{eqnarray}
\phi_0&\!=\!&0\,,\\
\phi_1&\!=\!&-(1/2) \arccos [ (\eta_1^2 - 2\eta_0^2)/(2\rho_0\eta_0)]\,,\\
\phi_2&\!=\!& \arccos[-\eta_1^2/(2\eta_2\eta_0)]. \label{eq:phi2}
\end{eqnarray}
We emphasize that the phases $\phi_1$ and $\phi_2$ are functions
of the diffraction efficiencies and therefore vary depending on
the properties of the grating.
This contrasts to the properties of mirrors, where the phase shift
between transmitted and reflected beams is independent of the
transmittance and reflectance coefficients.
Since the phase $\phi_2$ is a real number, the modulus of the
argument of the $\arccos$ in equation~(\ref{eq:phi2}) must be
smaller or equal to one and the following upper and lower limits
for $\eta_0$ and $\eta_2$ for a given reflectivity $\rho_0$ can be
derived, namely
\begin{equation}\label{eq:limits}
\eta_{0,\mathrm{_{min}^{max}}}=\eta_{2,\mathrm{_{min}^{max}}}=(1\pm\rho_0)/2.
\end{equation}
It should be noted that these limits are fundamental in the sense
that a reflection grating can only be designed and manufactured
having diffraction efficiencies within these boundaries. Equations
(\ref{eq:energycon})~-~(\ref{eq:limits}) provide a full set of
3-port coupling relations.
\begin{figure}[h]\centerline{\scalebox{0.95}{\includegraphics{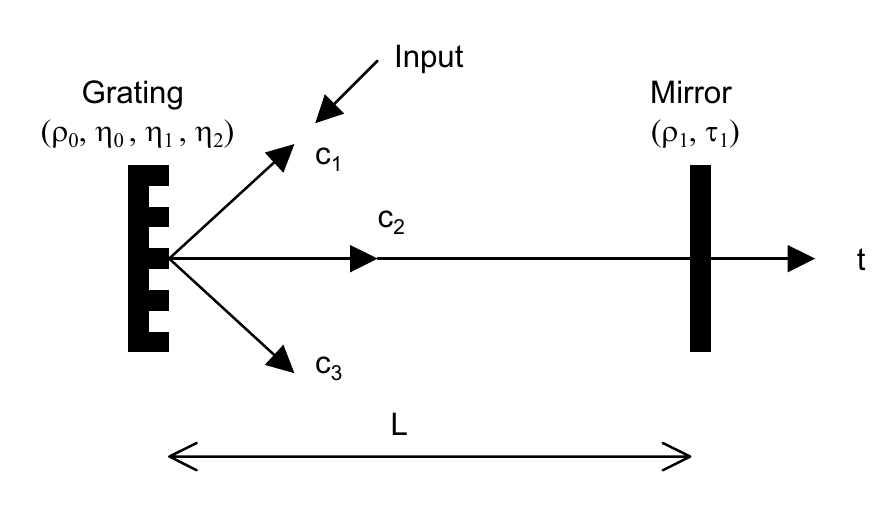}}}
  \caption{A Fabry-Perot cavity with a 3-port grating coupler and
  a conventional end mirror. The amplitudes of the fields of interest
  $(c_1, c_2, c_3, t)$ are indicated by arrows.}
   \label{fig:cavityschema}
\end{figure}

Knowledge of the scattering matrix $\bm{S}$ in Eq. (\ref{eq:S})
enables the calculation of input-output relations of
interferometric topologies. Here we consider the 3-port grating
coupled Fabry-Perot cavity. The  grating cavity is formed by
placing a mirror with amplitude reflectivity $\rho_1$  at a
distance $L$ parallel to the grating surface as it is illustrated
in Fig. \ref{fig:cavityschema}. To characterize the cavity, the
amplitudes $c_1, c_3$ for the two waves reflected from the cavity
and the intra-cavity amplitude $c_{2}$ are calculated as a
function of the cavity length.
% folgende Zeilen wurden wegen Platzmangel gelöscht
%\color{red} The amplitude $c_1$  of the reflection back  to the
%laser  as well as the additional reflectance $c_3$ are due to two
%interfering waves, namely the light coming from the cavity and the
%2nd and 0th order diffraction of the input light at the grating,
%respectively. We emphasize the difference to a conventional cavity
%which only has one reflection port where two waves interfere.
\color{black} Assuming unity input and no input at port 3 the
cavity is described by
\begin{equation}
\left(%
\begin{array}{c}
  c_1 \\
  c_2 \\
  c_3 \\
\end{array}%
\right) = \bm{S}_{3p}\times
\left(%
\begin{array}{c}
  1 \\
  \rho_1 c_2\exp({2i\phi}) \\
  0 \\
\end{array}%
\right).
\end{equation}
Solving for the amplitudes yields
\begin{eqnarray}\label{eq:amplitudes}
c_1&=& \eta_2\exp(i\phi_2) + \eta_1^2\exp[2i(\phi_1+\phi)]d,\\
c_2&=& \eta_1 \exp(i\phi_1)d,\\\label{eq:amplitude_c3}
c_3&=&\eta_0 + \eta_1^2\exp[2i(\phi_1+\phi)]d\,,\\
t&=&i\tau_1 c_2 \exp(i\phi)\,.
\end{eqnarray}
where $\phi = \omega L/c$ is the tuning parameter, $d$ is given
according to equation~(\ref{eq:d}), and $t$ is the amplitude of
the light transmitted through the cavity. The light power at the
different ports is proportional to the squared moduli  of the
amplitudes. The power gain inside the cavity is given by $|c_2|^2=
|\eta_1 d|^2$ analogous to equation (\ref{eq:gain}) for a
conventional cavity. In contrast to the power gain the power in
the two reflecting ports $|c_1|^2$ and $|c_3|^2$ depend on
$\eta_2$ and $\eta_0$.
\begin{figure}[h]\centerline{\scalebox{1}{\includegraphics{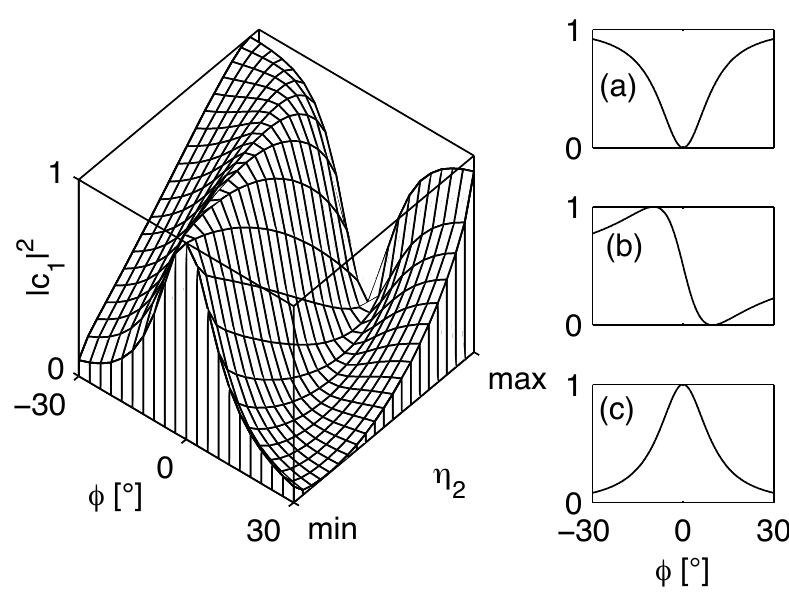}}}
  \caption{Power $|c_1|^2$of cavity back reflecting port  for gratings
  of different values of $\eta_2$. Left: Power as a function of
  $\phi$ and $\eta_2$. Right: Power as a function of $\phi$ for
  (a) $\eta_2=\eta_{2,\mathrm{max}}$; (b) $\eta_2=
  [(\eta_{2,\mathrm{max}}^2+\eta_{2,\mathrm{min}}^2)/2]^{1/2}; $ (c)
  $\eta_2=\eta_{2,\mathrm{min}}$. Cavity parameters: $\rho_0^2=0.5$,
  $\rho_1=1$.}
  \label{fig:etaminmax}
\end{figure}
Fig. \ref{fig:etaminmax} illustrates how the power out of the back
reflecting port varies as a function of $\eta_2$ and the tuning
$\phi$ of the cavity. For simplicity a cavity with a perfect end
mirror $\rho_1=1$ is assumed. For a coupler with $\eta_2=
\eta_{2,\mathrm{max}}$, the cavity does not reflect any light back
to the laser for a tuning of $\phi=0$. This corresponds to an
impedance matched cavity that transmits all the light on
resonance. For a coupler with $\eta_{2,\mathrm{min}}$, the
situation is reversed and all the light is reflected back to the
laser. For all other values of $\eta_2$ the back-reflected
intensity has intermediate values and as a significant difference
to conventional cavities: the intensity as a function of
cavity-tuning is no longer symmetric to the $\phi=0$ axis.

\begin{figure}[h]\centerline{\scalebox{1}{\includegraphics{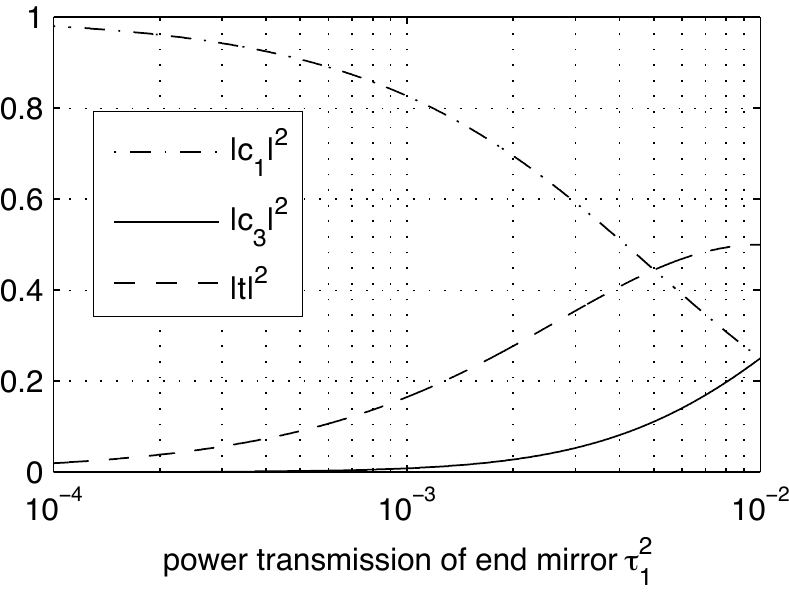}}}
  \caption{Powers of the two reflected ports and the transmitting
  port as a function of end mirror transmittance $\tau_1^2$ for a
  coupler with $\rho_0^2=0.99$ and $\eta_2=\eta_{2,\mathrm{min}}$
  for a tuning of $\phi=0$. }
  \label{fig:lossinfluence}
\end{figure}

Finally, we investigate the influence of loss in the cavity for a
coupler with $\eta_{2,\mathrm{min}}$. Fig.~\ref{fig:lossinfluence}
illustrates the effect of an end mirror with transmittance
$\tau_1>0$ to the power of the two reflecting ports of the cavity
on resonance. As a result, apart from the intra-cavity field,
losses affect mainly the back-reflecting port (dashed dotted
line). The effect on the dark port (solid line) is minor as it
stays essentially dark as long as the loss $\tau_1^2$ is small
compared to the coupling $\eta_1^2$.

In conclusion, we have investigated the three-port  reflection
grating and have derived its coupling relations. A three-port
device can be used to couple light into a Fabry-Perot cavity. The
input output relations of such a 3-port coupled cavity have
revealed substantial differences from a conventional cavity. A
grating with minimal $\eta_2$ is suitable for a coupler to an arm
cavity (single ended cavity) of a gravitational wave Michelson
interferometer. On resonance all power is reflected back to the
beam splitter of the interferometer. Hence no power is lost to the
additional port. This enables power recycling which is used in all
first and probably also in second and third generation detectors.
Furthermore we can calculate the phase signals carried by the
fields in equations~(\ref{eq:amplitudes}) and
(\ref{eq:amplitude_c3}) when changing the cavity length L and find
that the additional port splits a cavity strain signal. However,
the complete strain signal is still accessible to detection. This
will be the subject of a more detailed investigation in an
upcoming paper.

This research was supported by the Deutsche Forschungsgemeinschaft
within the Sonderforschungsbereich TR7. We thank P.~Beyersdorf,
T.~Clausnitzer, E.-B.~Kley, R.~Schilling, and B.~Willke for
helpful discussions. A.~Bunkowski's e-mail address is
alexander.bunkowski@aei.mpg.de

\end{document}